# Superparticle in an external chiral matter superfield

**Yu.A. Markov, M.A. Markova, A.I. Bondarenko, N.Yu. Markov**

*Matrosov Institute for System Dynamics and Control Theory, Siberian Branch, Russian Academy of Sciences, Irkutsk, 664033 Russia*

*E-mail:* markov@icc.ru, markova@icc.ru, 370omega@mail.ru, karvedys1398@mail.ru

Abstract: In this paper the interaction Lagrangian of a superparticle with an external chiral (matter) superfield $\Phi(x, \theta, \bar{\theta})$, invariant with respect to the global supersymmetry, is proposed. The kinematics of the superparticle is defined in an extended superspace described by superspace coordinates $(x_\mu, \theta_\alpha, \bar{\theta}_{\dot\alpha})$, where $x_\mu$ are pure bosonic coordinates, $\mu = 0, ..., 3$; $\theta_\alpha$ and $\bar{\theta}_{\dot\alpha}$ are additional fermionic Grassmann-valued spinors. With the purpose of the construction of the required Lagrangian, the chiral analogues of the supervector $A_\mu(x, \theta, \bar{\theta})$ and superspinor $A_\alpha(x, \theta, \bar{\theta})$, $\bar{A}_{\dot\alpha}(x, \theta, \bar{\theta})$ gauge fields are introduced, which are also matrices in color space. The problem considered in this paper is the further step in deriving the dynamical equations of motion for a spin color-charged particle moving in an external fermionic matter field. These equations of motion are of fundamental importance, for example, when taking into account the influence of fermionic background fields induced in a hot quark-gluon plasma (QGP) by thermal fluctuations on the dynamics of hard color-charged partons crossing the QGP. In turn, the solution of this problem will make it possible, at least at the qualitative level, to advance in the study of the motion in the external fermion matter field of such a complex physical object as a string.

# Contents

# Introduction

The problem of generalization of the well-known Brink-Schwartz Lagrangian for a massless free N = 1 superparticle [1] to the case of an external supersymmetric gauge field, apparently, was first discussed in [2]. The last work was devoted to the study of the classical superstring mechanics in the Hamiltonian and Lagrangian formulations in the presence of the massless external fields. Here the superparticle was considered as a certain limiting case of this mechanics. Siegel in [2] proposed to modify the standard Brink-Schwartz expression by adding to it two terms with Lagrange multipliers, one of which obviously takes into account the massless of the superparticle and the second term takes into account the presence of an additional fermionic constraint in the system. The interaction with an external super-Yang-Mills field was introduced by covariantizing the vector $\partial_\mu$ and spinor $D_\alpha$ derivatives with vector and spinor gauge superpotentials $A_\mu(x,\theta,\bar{\theta})$ and $A_\alpha(x,\theta,\bar{\theta})$, correspondingly.

Further, in [3] the most general classical action for a massless superparticle moving in an external N = 1 super-Maxwell field has been proposed and corresponding classical equations of motion have been written out. The previous paper is closely related to the work [4], in which the model of N = 1, D = 4 superparticles in the presence of N=1 supersymmetric Maxwell field uniformly filling Minkowski space-time is considered.

In a number of works (see below) a rather original approach to the construction of Lagrangians of free and interacting superparticles, which can be denoted as the superfield approach, has been proposed. It appeared in connection with the investigation of the problem of the exact relation between quantum dynamics of a particle moving along a superspace trajectory and superfield theories formulated within the usual second-quantized formalism. The main thesis, which has been presented there, is that the action for a massless (massive) superparticle should be the first-quantized equivalent of the conventional second-quantized supersymmetric action.

The first work in this direction was the work [5]. The initial action in this work was the action for a massless vector superfield with covariant ξ-gauge. Using the method originally proposed by R. Feynman [6], a first-quantized propagator for the vector superfield was found from this action in the form of integration over trajectories in superspace.

In another, also interesting and important paper [7], the reparametrization-invariant action for a superparticle was constructed based on the representations of the one-particle Green's function in the form of an integral over trajectories in superspace, but not for a vector, but for a chiral superfield interacting with an external super-Maxwell field. The ideas in these two papers have been developed in a number of other papers [8-10]. In [11], the Brink-Schwartz superparticle interacting with D = 4, N = 1 supergravity was considered within the superfield approach.

Here the question arises whether, within only the superfield approach, it is possible to define the classical action for a superparticle moving in an external chiral (matter) superfield in the spirit of [8], where the superparticle moving in an external vector (Maxwellian) superfield was considered. Most likely, here it is necessary to use the action for the massless vector superfield as the basic initial action, as it was done in [5], however, adding interaction terms with the chiral field. It is necessary to take the latter as an external superfield and to find a representation in the form of a path integral in superspace for the vector superfield propagator, in the presence of a given external chiral superfield in the system. Due to essential nonlinearity, here, most likely, it will be necessary to work in the Wess-Zumin gauge for the vector superfield.

In constructing the interaction Lagrangian of a spin superparticle with an external fermionic matter superfield, we inevitably encounter the necessity to introduce into consideration the concept of a spinor superfield. It is known [12, 13] that spinor superfields, as well as tensor superfields of arbitrary rank, represent a new category of local representations of the supersymmetry algebra and are thus a natural generalization of scalar superfields. With respect to translation and supertranslation these fields are transformed as scalars, but with a homogeneous (intrinsic) Lorentz group they are transformed as spinors and tensors. Such generalizations are applicable to both chiral and nonchiral superfields. In particular, a vector multiplet can be expressed through a chiral spinor superfield playing the role of a generalized supersymmetric stress tensor.

External supersymmetric gauge fields, although they do not represent (fermionic) matter superfields, nevertheless, contain in their field decomposition the fermionic components and thus already give some possibility to analyze the dynamics of the (spin) superparticle interacting with a fermionic component of the SUSY gauge field. A noticeable simplification here will be, for example, in the analysis of the color degree dynamics of the superparticle, due to the fact that both the bosonic and fermionic components of the external gauge superfield lie in the same adjoint representation of a gauge group. The latter circumstance requires introducing into consideration only one set of additional fermionic functions to describe the color degree of freedom of the superparticle.

In order to obtain the action of a spin color-charged superparticle moving in an external fermionic (chiral, matter) superfield as a first step one can try to use already known actions for a superparticle in external supersymmetric gauge abelian or non-Abelian fields. It is known that the gauge superpotentials $A_\mu(x,\theta,\bar\theta)$, $A_\alpha(x,\theta,\bar\theta)$, $A_{\dot\alpha}(x,\theta,\bar\theta)$, which enter into the interaction Lagrangian of a superparticle with an external gauge superfield can be represented in the form of the gauge- and supersymmetric-covariant derivatives of some scalar superfields $V(x,\theta,\bar\theta)$ and $U(x,\theta,\bar\theta)$ [14]. On these scalar superfields no additional constraints are imposed, so they are called *prepotentials*. The canonical dimension of these prepotentials (in mass units) is zero. Further, it is well known that any chiral superfield $\Phi(x,\theta,\bar\theta)$ can also be represented in the form of gauge- and supersymmetric-covariant derivatives acting on some scalar superfield $S(x,\theta,\bar\theta)$ [14,15]. This scalar superfield also does not obey any additional conditions and its canonical dimension is zero. In fact, it plays the role of a prepotential for the chiral matter superfield and provides an alternative description of the latter. The initial chiral superfield $\Phi$ itself can be considered as a chiral stress tensor for the "gauge" superfield $S(x,\theta,\bar\theta)$. The chiral prepotential $S(x,\theta,\bar\theta)$ can be expressed (nonlocally) in terms of the initial chiral superfield $\Phi(x,\theta,\bar\theta)$.

Because of the great similarity of properties of gauge and chiral prepotentials, the first simple and most natural step in constructing the interaction Lagrangian of a superparticle with an external chiral matter superfield is to replace the gauge prepotentials $V(x,\theta,\bar\theta)$ and $U(x,\theta,\bar\theta)$ in the well-known interaction Lagrangian of a superparticle with the super-Yang-Mills (super-Maxwell) field by prepotentials $S(x,\theta,\bar\theta)$ and $\bar S(x,\theta,\bar\theta)$ corresponding (anti)chiral superfields.

Since we know the relation between the prepotential $S(x,\theta,\bar\theta)$ and the initial matter superfield $\Phi(x,\theta,\bar\theta)$, we can immediately write out a model Lagrangian of interaction of the superparticle with the component fields included in the chiral supermultiplet. The disadvantage of such "direct" approach is a possible nonlocality of some interaction terms.

On the other hand, this approach naturally introduces an interaction with the external gauge superfield by simply replacing the SUSY covariant derivatives by supersymmetric gauge-covariant derivatives. In this case the inverse expression of the chiral prepotential in terms of the chiral-covariant superfield $\Phi(x,\theta,\bar{\theta})$ will be some nonlocal function depending on the external gauge superpotentials.

Finally, we note that the approach presented here is essentially a further development and generalization of the ideas presented in our earlier work [16] on constructing the action describing dynamics of a classical color-charged particle interacting with background non-Abelian gauge and fermion fields. This problem is closely related to the study of the scattering processes of hard color-charged particles off soft gluon and quark–antiquark fields. These soft fields are induced by thermal fluctuations in a hot quark–gluon plasma [17, 18].

## 1 Spinning superparticle in external super-symmetric gauge and chiral fields

Let us generalize the notion of an ordinary bosonic point particle moving in Minkowski space to a particle moving in superspace. As an initial expression we take the following simplest action:

$$S = -\tfrac{1}{2}\int \left(\frac{\dot{x}^2}{e} + em^2\right) d\tau, \quad (1.1)$$

where $e$ is the (one-dimensional) vierbein field.

We consider first the massless case, $m = 0$. One can write several different actions, since we have two expressions:

$$\dot{x}^\mu - i\bar{\theta}^I \Gamma^\mu \dot{\theta}^I,$$
$$\dot{\theta}^I_\alpha, \quad (1.2)$$

which are invariant with respect to the global supersymmetry. Lorentz-invariant Lagrangians of different kind can be constructed from them. Let us consider a simple variant, which directly generalizes the bosonic action (1.1), using the first invariant of (1.2):

$$S = -\frac{1}{2}\int \frac{1}{e}\left(\dot{x}^\mu - i\bar{\theta}^I\Gamma^\mu\dot{\theta}^I\right)^2 d\tau. \qquad (1.3)$$

Here, we come to the Brink-Schwartz superparticle [1]. This expression possesses obvious Lorentz invariance and supersymmetry, and hence super Poincaré invariance. From the action (1.3) it follows the equations of motion:

$$\begin{aligned}
\delta e: &\quad p_\mu p^\mu = 0, \quad p^\mu \equiv \dot{x}^\mu - i\bar{\theta}^I\Gamma^\mu\dot{\theta}^I, \\
\delta x_\mu: &\quad \dot{p}_\mu = 0, \\
\delta\bar{\theta}^I: &\quad \left(\Gamma^\mu p_\mu\right)\dot{\theta}^I = 0.
\end{aligned} \qquad (1.4)$$

## 1.1 The Weyl representation. Superparticle in an external gauge field

Let us take a closer look at the action of a free massless superparticle (1.3). For the specificity we consider the case D = 4, i.e., we set $\Gamma^\mu \equiv \gamma^\mu$. Next, we restrict our consideration to the number of supersymmetries N = 1, and set the spinor $\theta_\alpha$ to be a Majorana one. In this case, it is more convenient to pass to the two-component formalism of the Weyl representation, i.e., the original spinor $\theta_\alpha$, where $\alpha = 1,\ldots,4$, is represented in the following form:

$$\theta = (\theta_\alpha) = \begin{pmatrix} \theta_\alpha \\ \bar{\theta}^{\dot\alpha} \end{pmatrix}, \quad \bar{\theta} = (\bar{\theta}_{\dot\alpha}) = (\theta^\alpha\,\bar{\theta}_{\dot\alpha}),$$
$$\bar{\theta}_{\dot\alpha} = (\theta_\alpha)^*, \quad \alpha, \dot\alpha = 1,2 \qquad (1.5)$$

and the Dirac gamma matrices take the form:

$$\gamma^\mu = \begin{pmatrix} 0 & \sigma^\mu \\ \bar{\sigma}^\mu & 0 \end{pmatrix},$$
$$\sigma^\mu = (I, \boldsymbol{\sigma}) = \left(\sigma^\mu\right)_{\alpha\dot\alpha},$$
$$\bar{\sigma}^\mu = (I, -\boldsymbol{\sigma}) = \left(\bar{\sigma}^\mu\right)^{\dot\alpha\alpha}. \qquad (1.6)$$

Further, we need only one property for (anticommutative) two-component spinors:

$$\bar{\psi}\bar{\sigma}^\mu\chi = -\chi\sigma^\mu\bar{\psi}. \qquad (1.7)$$

Thus, for $N = 1$ supersymmetry we have, within the framework of this two-component formalism, a superspace $(x^\mu, \theta_\alpha, \bar{\theta}^{\dot\alpha})$. In the original action (1.3) we describe in more detail the contribution with spinors:

$$\bar{\theta}\gamma^\mu\dot{\theta} = \begin{pmatrix}\theta^\alpha & \bar{\theta}_{\dot{\alpha}}\end{pmatrix}\begin{pmatrix} 0 & (\sigma^\mu)_{\alpha\dot{\alpha}} \\ (\bar{\sigma}^\mu)_{\alpha\dot{\alpha}} & 0 \end{pmatrix}\begin{pmatrix}\dot{\theta}_\alpha \\ \dot{\bar{\theta}}^{\dot{\alpha}}\end{pmatrix} =$$
$$= \bar{\theta}\bar{\sigma}^\mu\dot{\theta} + \theta\sigma^\mu\dot{\bar{\theta}} = -\dot{\theta}\sigma^\mu\bar{\theta} + \theta\sigma^\mu\dot{\bar{\theta}}, \qquad (1.8)$$

and the action itself in two-component notation takes the form:

$$S = -\frac{1}{2}\int \frac{1}{e}\left(\dot{x}^\mu(\tau) - i\left[\left(\theta\sigma^\mu\dot{\bar{\theta}}\right) - \left(\dot{\theta}\sigma^\mu\bar{\theta}\right)\right]\right)^2 d\tau. \qquad (1.9)$$

The supersymmetry transformations with infinitesimal spinor parameter $\varepsilon_\alpha$:

$$\delta x_\mu = i[(\varepsilon\sigma^\mu\bar{\theta}) - (\theta\sigma^\mu\bar{\varepsilon})], \quad \delta\theta_\alpha = \varepsilon_\alpha, \quad \delta\bar{\theta}^{\dot{\alpha}} = \bar{\varepsilon}^{\dot{\alpha}} \qquad (1.10)$$

leave the action (1.9) invariant.

The interaction term of a charged spinless particle with an external gauge field has the following form:

$$S = -q \int A_\mu(x)\dot{x}^\mu(\tau)d\tau. \qquad (1.11)$$

The simplest and most natural way to generalize this expression to the case of a charged superparticle is to "lengthen" the 4-speed

$$\dot{x}^\mu \to \dot{x}^\mu - i\left[\left(\theta\sigma^\mu\dot{\bar{\theta}}\right) - \left(\dot{\theta}\sigma^\mu\bar{\theta}\right)\right]. \qquad (1.12)$$

Next, it is necessary to generalize the four-potential $A_\mu(x)$, i.e., to write

$$A_\mu(x) \to \mathcal{A}_\mu(x,\theta,\bar{\theta}). \qquad (1.13)$$

Thus, the gauge potential becomes a "true" superfield, as it is the case in supersymmetry theory, which in addition has an external vector index.

## 1.2 Superparticle in super-Maxwell and -Yang-Mills external fields

We consider the following expression as the interaction term for a charged superparticle with an abelian N = 1 supersymmetric gauge field:

$$S_{int} = -q\int\left(\dot{x}^\mu - i\left[\left(\theta\sigma^\mu\dot{\bar{\theta}}\right) - \left(\dot{\theta}\sigma^\mu\bar{\theta}\right)\right]\right)\mathcal{A}_\mu(x,\theta,\bar{\theta})d\tau. \qquad (1.14)$$

By virtue of the nilpotency of the Grassmann coordinates $\theta_\alpha$ and $\bar{\theta}_{\dot{\alpha}}$ the superfield in (1.14) can be represented as a series expansion in these spinor variables:

$$\mathcal{A}_\mu(x,\theta,\bar{\theta}) = A_\mu(x) + \theta^\alpha \psi_{\mu\alpha}(x) + \bar{\theta}_{\dot\alpha}\bar{\psi}_\mu^{\dot\alpha}(x) + \dots . \qquad (1.15)$$

To clarify the physical meaning of this superfield and to determine the possible additional contributions to $S_{\text{int}}$ it is necessary to turn to the theory of supersymmetric gauge fields.

Back then, J. Wess and B. Zumino proposed to use differential geometry in superspace, which allowed one to better understand supersymmetric gauge theory. This approach has the advantage that it is the closest to the construction of the usual (non)Abelian gauge theory. In supersymmetric theory (recall that we consider the case of $N = 1$ SUSY) we have covariant superspace derivatives:

$$\begin{aligned}
\partial_\mu &\equiv \frac{\partial}{\partial x^\mu}, \\
D_\alpha &\equiv \frac{\partial}{\partial \theta^\alpha} + i\sigma^\mu_{\alpha\dot\alpha}\bar{\theta}^{\dot\alpha}\frac{\partial}{\partial x^\mu}, \\
\bar{D}_{\dot\alpha} &\equiv -\frac{\partial}{\partial \bar{\theta}^{\dot\alpha}} - i\theta^\alpha \sigma^\mu_{\alpha\dot\alpha}\frac{\partial}{\partial x^\mu}.
\end{aligned} \qquad (1.16)$$

For each superspace derivative, we introduce the corresponding gauge potential superfields

$$\mathcal{A}_\mu(x,\theta,\bar{\theta}), \quad \mathcal{A}_\alpha(x,\theta,\bar{\theta}), \quad \bar{\mathcal{A}}_{\dot\alpha}(x,\theta,\bar{\theta}). \qquad (1.17)$$

In principle, it is possible to consider that $\mathcal{A}_\mu \neq (\mathcal{A}_\mu)^\dagger$ and $\bar{\mathcal{A}}_{\dot\alpha} \neq (\mathcal{A}_\alpha)^\dagger$. The Yang-Mills field potentials are superfields taking values in a Lie algebra:

$$\mathcal{A}_\mu(x,\theta,\bar{\theta}) = \sum_a t^a \mathcal{A}_\mu^a(x,\theta,\bar{\theta}), \quad \mathcal{A}_\alpha(x,\theta,\bar{\theta}) = \sum_a t^a \mathcal{A}_\alpha^a(x,\theta,\bar{\theta}), \dots . \qquad (1.18)$$

## 1.3 Superparticle in an external field N = 1 super-Maxwell

In the previous section we have written out the expression of the form (1.14) as a term for the interaction of a superparticle with an abelian external superfield $\mathcal{A}_\mu$. However, there exist two other potentials with spinor indices: $\mathcal{A}_\alpha(x,\theta,\bar{\theta})$ and $\bar{\mathcal{A}}_{\dot\alpha}(x,\theta,\bar{\theta})$, which must also enter into the total interaction Lagrangian. In constructing a supersymmetric Lagrangian for a free superparticle, we considered three types of SUSY invariants:

$$\dot{x}^\mu - i\left[\theta\sigma^\mu\dot{\bar{\theta}} - \dot{\theta}\sigma^\mu\bar{\theta}\right], \quad \dot{\theta}_\alpha \quad \text{and} \quad \dot{\bar{\theta}}_{\dot\alpha}. \qquad (1.19)$$

We have used only the first of them. The other invariants can be employed to form the other interaction terms of the type $\dot{\theta}^\alpha A_\alpha$ and $\dot{\bar{\theta}}_{\dot{\alpha}}\bar{A}^{\dot{\alpha}}$. Thus, we have the following interaction action for the superparticle in an external Maxwell supersymmetric field:

$$S_{int} = -q\int\left[\left(\dot{x}^\mu - i\left[\left(\theta\sigma^\mu\dot{\bar{\theta}}\right) - \left(\dot{\theta}\sigma^\mu\bar{\theta}\right)\right]\right)A_\mu(x,\theta,\bar{\theta}) + \right.$$
$$\left. + i\dot{\theta}^\alpha A_\alpha(x,\theta,\bar{\theta}) - i\dot{\bar{\theta}}_{\dot{\alpha}}\bar{A}^{\dot{\alpha}}(x,\theta,\bar{\theta})\right]d\tau. \quad (1.20)$$

In quantization of the model with the complete action $S_{tot} = S_0 + S_{int}$, the relations on the $U(1)$ stress tensors automatically arise, by virtue of which the structure of the abelian gauge superfield potentials is fixed:

$$A_\alpha = -iD_\alpha V,$$
$$\bar{A}_{\dot{\alpha}} = i\bar{D}_{\dot{\alpha}}U, \quad (1.21)$$
$$A_\mu = \frac{i}{4}\bar{\sigma}_\mu^{\dot{\beta}\alpha}(D_\alpha\bar{A}_{\dot{\beta}} + \bar{D}_{\dot{\beta}}A_\alpha).$$

The reality condition of the abelian gauge superfield potentials determines the relationship between the prepotentials $V$, $U$:

$$V + U = (V + U)^\dagger. \quad (1.22)$$

The expression $S_{int}$ is invariant with respect to the space-time SUSY (1.10) and gauge transformations $\delta A_\mu = \partial_\mu\Lambda$, $\delta A_\alpha = D_\alpha\Lambda$, ... .

## 1.4 Superparticle in an external chiral superfield

So far, we have considered the motion of a superparticle in an external supersymmetric gauge field, which itself is the carrier of interaction between matter particles. Then the question arises whether it is possible to consider the interaction of a superparticle with an external chiral matter superfield $\Phi(x,\theta,\bar{\theta})$. Let us recall that in the case of the super-Maxwell field the potential $A_\alpha(x,\theta,\bar{\theta})$ is expressed through the prepotential $V(x,\theta,\bar{\theta})$ as follows:

$$A_\alpha(x,\theta,\bar{\theta}) = -iD_\alpha V(x,\theta,\bar{\theta}),$$

where $V$ – is an arbitrary superfield. The chiral superfield $\Phi(x,\theta,\bar{\theta})$ satisfying the equality

$$\bar{D}_{\dot{\alpha}}\Phi(x,\theta,\bar{\theta}) = 0, \tag{1.23}$$

can in turn be expressed through the nonchiral potential superfield $S(x,\theta,\bar{\theta})$ [13,14]:

$$\Phi = \bar{D}^2 S = \bar{D}_{\dot{\alpha}} \bar{D}^{\dot{\alpha}} S,$$
$$\bar{\Phi} = D^2 S^\dagger = D^\alpha D_\alpha S^\dagger, \tag{1.24}$$

by virtue of the fact that $\bar{D}^3 = D^3 = 0$. This superfield $S$ plays the role of a "prepotential" for the chiral superfield. The superfield $S$ can be expressed in terms of the initial chiral field $\Phi$:

$$S(x,\theta,\bar{\theta}) = \frac{1}{16}\frac{1}{\Box} D^2 \Phi(x,\theta,\bar{\theta}). \tag{1.25}$$

The connection, however, is non-local.

The analog of the gauge transformation $A_\mu \to A_\mu + \partial_\mu \Lambda$ here is a transformation of the form $S \to S + \bar{D}X$, where $X$ is an arbitrary superfield playing the role of the gauge parameter $\Lambda$. The canonical dimension of the field and the derivatives are, respectively $[S] = 0$ and $[D] = [\bar{D}] = \frac{1}{2}$. Let us define the interaction with the external chiral (matter) superfield simply setting by analogy

$$\dot{\theta}^\alpha A_\alpha(x,\theta,\bar{\theta}) = -i\dot{\theta}^\alpha D_\alpha V(x,\theta,\bar{\theta}) \;\Rightarrow\; -i\dot{\theta}^\alpha D_\alpha S(x,\theta,\bar{\theta}), \tag{1.26}$$

$$\dot{\bar{\theta}}_{\dot{\alpha}} \bar{A}^{\dot{\alpha}}(x,\theta,\bar{\theta}) = i\dot{\bar{\theta}}_{\dot{\alpha}} \bar{D}^{\dot{\alpha}} U(x,\theta,\bar{\theta}) \;\Rightarrow\; i\dot{\bar{\theta}}_{\dot{\alpha}} \bar{D}^{\dot{\alpha}} S^\dagger(x,\theta,\bar{\theta}), \tag{1.27}$$

$$\dot{x}_\mu A^\mu(x,\theta,\bar{\theta}) = -\frac{1}{4}\dot{x}_\mu \left(\bar{\sigma}^\mu\right)^{\dot{\beta}\alpha} (D_\alpha \bar{D}_{\dot{\beta}} U - \bar{D}_{\dot{\beta}} D_\alpha V) \;\Rightarrow$$
$$\Rightarrow\; \frac{1}{4}\dot{x}_\mu \left(\bar{\sigma}^\mu\right)^{\dot{\beta}\alpha} (\bar{D}_{\dot{\beta}} D_\alpha S - D_\alpha \bar{D}_{\dot{\beta}} S^\dagger). \tag{1.28}$$

Then the analog the interaction Lagrangian in the action (1.20) will be the Lagrangian:

$$L_{int} = q\left[\frac{1}{4}\left(\dot{x}^\mu - i\left[\left(\theta\sigma^\mu\dot{\bar{\theta}}\right) - \left(\dot{\theta}\sigma^\mu\bar{\theta}\right)\right]\right)\left(\bar{\sigma}_\mu\right)^{\dot{\beta}\alpha} (\bar{D}_{\dot{\beta}} D_\alpha S - D_\alpha \bar{D}_{\dot{\beta}} S^\dagger) - \right.$$
$$\left. -\dot{\theta}^\alpha D_\alpha S(x,\theta,\bar{\theta}) + \dot{\bar{\theta}}_{\dot{\alpha}} \bar{D}^{\dot{\alpha}} S^\dagger(x,\theta,\bar{\theta})\right]. \tag{1.29}$$

Our task now is to obtain an explicit form of (1.29) based on knowledge of the structure of the chiral superfield $\Phi(x,\theta,\bar{\theta})$ and of the covariant spinor derivatives $D$ and $\bar{D}$.

# 2 The interaction Lagrangian of a superparticle with an external chiral matter superfield

## 2.1 The fourth-order covariant derivative of chiral superfield $\Phi(x,\theta,\bar\theta)$

Recall that the chiral superfield obeys the condition

$$\bar D_{\dot\alpha}\Phi(x,\theta,\bar\theta) = 0. \tag{2.1a}$$

Its most general solution has the form

$$\Phi(x,\theta,\bar\theta) = A(x) + i\theta\sigma^\eta\bar\theta\partial_\eta A(x) + \tfrac{1}{4}\theta\theta\bar\theta\bar\theta\Box A(x) + \sqrt{2}\theta\psi(x) - \tfrac{i}{\sqrt{2}}\theta\theta\partial_\eta\psi(x)\sigma^\eta\bar\theta +$$
$$+ \theta\theta F(x). \tag{2.1b}$$

Let us calculate the fourth-order derivative of a chiral superfield $\Phi(x,\theta,\bar\theta)$ of the form

$$(\overline{DD})(DD)\Phi(x,\theta,\bar\theta). \tag{2.2}$$

For this purpose, we need to prove the following relation for the commutator of covariant spinor derivatives $D$ and $\bar D$ [16]:

$$[DD,\bar D\bar D] = -16\Box + 8i\partial_\nu D\sigma^\nu\bar D. \tag{2.3}$$

Calculating the commutator, making use of the equality [15]

$$[D_\alpha,\bar D^2] = -4i\sigma_{\nu\alpha\dot\beta}\partial^\nu\bar D^{\dot\beta}, \tag{2.4}$$

we obtain

$$[D^\alpha D_\alpha,\bar D^2] = D^\alpha[D_\alpha,\bar D^2] + [D^\alpha,\bar D^2]D_\alpha = D^\alpha[D_\alpha,\bar D^2] - [D_\alpha,\bar D^2]D^\alpha =$$
$$= -4i\sigma_{\mu\alpha\dot\beta}\partial^\mu D^\alpha\bar D^{\dot\beta} + 4i\sigma_{\mu\alpha\dot\beta}\partial^\mu\bar D^{\dot\beta}D^\alpha =$$
$$= -4i\sigma_{\mu\alpha\dot\beta}\partial^\mu[D^\alpha,\bar D^{\dot\beta}].$$
$$\tag{2.5}$$

Next, the equality

$$[D_\alpha,\bar D_{\dot\beta}] = \sigma_{\nu\alpha\dot\beta}D\sigma^\nu\bar D + 2i\sigma_{\nu\alpha\dot\beta}\partial^\nu, \tag{2.6}$$

results (2.5) in the following form:

$$[D^\alpha D_\alpha,\bar D^2] = -4i\sigma_{\mu\alpha\dot\beta}\partial^\mu[D^\alpha,\bar D^{\dot\beta}] = -4i\varepsilon^{\dot\beta\dot\gamma}\varepsilon^{\alpha\beta}\sigma_{\mu\alpha\dot\beta}\partial^\mu(\sigma_{\nu\beta\dot\gamma}D\sigma^\nu\bar D + 2i\sigma_{\nu\beta\dot\gamma}\partial^\nu) =$$
$$= -4i\sigma_{\mu\alpha\dot\beta}\bar\sigma_\nu^{\dot\beta\alpha}\partial^\mu(D\sigma^\nu\bar D + 2i\partial^\nu) = -4i(-2g_{\mu\nu})(\partial^\mu D\sigma^\nu\bar D + 2i\partial^\mu\partial^\nu)$$
$$= 8i\partial_\nu D\sigma^\nu\bar D - 16\Box.$$
$$\tag{2.7}$$

As a consequence of the relation (2.3) we have

$$\left[DD, \bar{D}\bar{D}\right]\Phi(x,\theta,\bar{\theta}) = (D_\alpha D^\alpha)(\bar{D}_{\dot\beta}\bar{D}^{\dot\beta})\Phi(x,\theta,\bar{\theta}) - (\bar{D}_{\dot\beta}\bar{D}^{\dot\beta})(D_\alpha D^\alpha)\Phi(x,\theta,\bar{\theta}) =$$
$$= 8i\partial_\nu D\sigma^\nu \bar{D}\Phi - 16\Box\Phi(x,\theta,\bar{\theta}). \tag{2.8}$$

Ultimately, using the fact that $\Phi$ is a chiral superfield, i.e., it obeys the condition (2.1a) for the expression (2.2) we get:

$$(\bar{D}\bar{D})(DD)\Phi(x,\theta,\bar{\theta}) = 16\Box\Phi(x,\theta,\bar{\theta}). \tag{2.9}$$

The resulting formula (2.9) makes it possible to express the potential superfield $S(x,\theta,\bar{\theta})$ in terms of the chiral superfield $\Phi(x,\theta,\bar{\theta})$ (see (1.25)).

## 2.2. An explicit form of covariant spinor derivatives $D^\beta$ and $\bar{D}^{\dot\beta}$

We have the initial definitions of the superspace differential operators with the lower spinor indices

$$D_\alpha = \frac{\partial}{\partial\theta^\alpha} + i\sigma^\mu{}_{\alpha\dot\beta}\bar{\theta}^{\dot\beta}\partial_\mu, \quad \bar{D}_{\dot\alpha} = -\frac{\partial}{\partial\bar{\theta}^{\dot\alpha}} - i\theta^\beta \sigma^\mu{}_{\beta\dot\alpha}\partial_\mu. \tag{2.10}$$

Let us define similar operators with upper spinor indices $D^\beta$ and $\bar{D}^{\dot\beta}$, respectively:

$$D^\beta = \varepsilon^{\beta\alpha} D_\alpha = \varepsilon^{\beta\alpha}\frac{\partial}{\partial\theta^\alpha} + i\varepsilon^{\beta\alpha}\sigma^\mu{}_{\alpha\dot\beta}\bar{\theta}^{\dot\beta}\partial_\mu =$$
$$= -\frac{\partial}{\partial\theta_\beta} + i\varepsilon^{\dot\beta\dot\alpha}\varepsilon^{\beta\alpha}\sigma^\mu{}_{\alpha\dot\beta}\bar{\theta}_{\dot\alpha}\partial_\mu = -\frac{\partial}{\partial\theta_\beta} - i\varepsilon^{\dot\alpha\dot\beta}\varepsilon^{\beta\alpha}\sigma^\mu{}_{\alpha\dot\beta}\bar{\theta}_{\dot\alpha}\partial_\mu = \tag{2.11}$$
$$= -\frac{\partial}{\partial\theta_\beta} - i\bar{\theta}_{\dot\alpha}\bar{\sigma}^{\mu\dot\alpha\beta}\partial_\mu;$$

$$\bar{D}^{\dot\beta} = \varepsilon^{\dot\beta\dot\alpha}\bar{D}_{\dot\alpha} = -\varepsilon^{\dot\beta\dot\alpha}\frac{\partial}{\partial\bar{\theta}^{\dot\alpha}} - i\varepsilon^{\dot\beta\dot\alpha}\theta^\beta\sigma^\mu{}_{\beta\dot\alpha}\partial_\mu =$$
$$= \frac{\partial}{\partial\bar{\theta}_{\dot\beta}} - i\varepsilon^{\dot\beta\dot\alpha}\varepsilon^{\beta\alpha}\theta_\alpha\sigma^\mu{}_{\beta\dot\alpha}\partial_\mu = \frac{\partial}{\partial\bar{\theta}_{\dot\beta}} + i\varepsilon^{\dot\beta\dot\alpha}\varepsilon^{\alpha\beta}\sigma^\mu{}_{\beta\dot\alpha}\theta_\alpha\partial_\mu = \tag{2.12}$$
$$= \frac{\partial}{\partial\bar{\theta}_{\dot\beta}} + i\bar{\sigma}^{\mu\dot\beta\alpha}\theta_\alpha\partial_\mu.$$

Finally, we obtain an explicit form of covariant derivatives with upper spinor indices:

$$D^\beta = -\frac{\partial}{\partial\theta_\beta} - i\bar{\theta}_{\dot\alpha}\bar{\sigma}^{\mu\dot\alpha\beta}\partial_\mu, \quad \bar{D}^{\dot\beta} = \frac{\partial}{\partial\bar{\theta}_{\dot\beta}} + i\bar{\sigma}^{\mu\dot\beta\alpha}\theta_\alpha\partial_\mu. \tag{2.13}$$

## 2.3 Derivation of the expression for the derivative of the chiral superfield $D^2\Phi(x,\theta,\bar\theta)$

Our task now is to obtain an explicit form of the second-order covariant derivative of the chiral superfield $\Phi(x,\theta,\bar\theta)$, which enters the relation (1.25). The original expression is

$$D^\alpha D_\alpha \Phi(x,\theta,\bar\theta). \tag{2.14}$$

By making use of an explicit form of the superspace differential operators (2.10) and (2.13) and the decomposition of the chiral superfield $\Phi(x,\theta,\bar\theta)$, Eq. (2.1b), we have

$$D^\alpha D_\alpha \Phi(x,\theta,\bar\theta) = \left(-\frac{\partial}{\partial\theta_\alpha} - i\bar\theta_{\dot\beta}\bar\sigma^{\mu\dot\beta\alpha}\partial_\mu\right)\left(\frac{\partial}{\partial\theta^\alpha} + i\sigma^\nu{}_{\alpha\dot\gamma}\bar\theta^{\dot\gamma}\partial_\nu\right)\times$$

$$\times\left[A(x) + i\theta\sigma^\eta\bar\theta\partial_\eta A(x) + \frac{1}{4}\theta\theta\bar\theta\bar\theta\Box A(x) + \sqrt{2}\theta\psi(x) - \frac{i}{\sqrt{2}}\theta\theta\partial_\eta\psi(x)\sigma^\eta\bar\theta + \theta\theta F(x)\right].$$

$$\tag{2.15}$$

Consider the differentiation of the separate terms in $\Phi(x,\theta,\bar\theta)$ containing the component fields $\psi(x)$, $A(x)$ and $F(x)$:

$$D^\alpha D_\alpha \Phi^{(\psi)}(x,\theta,\bar\theta) = \left(-\frac{\partial}{\partial\theta_\alpha} - i\bar\theta_{\dot\beta}\bar\sigma^{\mu\dot\beta\alpha}\partial_\mu\right)\left(\frac{\partial}{\partial\theta^\alpha} + i\sigma^\nu{}_{\alpha\dot\gamma}\bar\theta^{\dot\gamma}\partial_\nu\right)\times$$

$$\times\left[\sqrt{2}\theta^\rho\psi_\rho(x) - \frac{i}{\sqrt{2}}\theta\theta\partial_\eta\psi(x)\sigma^\eta\bar\theta\right], \tag{2.16}$$

$$D^\alpha D_\alpha \Phi^{(A)}(x,\theta,\bar\theta) = \left(-\frac{\partial}{\partial\theta_\alpha} - i\bar\theta_{\dot\beta}\bar\sigma^{\mu\dot\beta\alpha}\partial_\mu\right)\left(\frac{\partial}{\partial\theta^\alpha} + i\sigma^\nu{}_{\alpha\dot\gamma}\bar\theta^{\dot\gamma}\partial_\nu\right)\times$$

$$\times\left[A(x) + i\theta\sigma^\eta\bar\theta\partial_\eta A(x) + \tfrac{1}{4}\theta\theta\bar\theta\bar\theta\Box A(x)\right], \tag{2.17}$$

$$D^\alpha D_\alpha \Phi^{(F)}(x,\theta,\bar\theta) = \left(-\frac{\partial}{\partial\theta_\alpha} - i\bar\theta_{\dot\beta}\bar\sigma^{\mu\dot\beta\alpha}\partial_\mu\right)\left(\frac{\partial}{\partial\theta^\alpha} + i\sigma^\nu{}_{\alpha\dot\gamma}\bar\theta^{\dot\gamma}\partial_\nu\right)\times$$

$$\times[\theta\theta F(x)]. \tag{2.18}$$

We have for the first expression (2.16)

$$D^\alpha D_\alpha \Phi^{(\psi)}(x,\theta,\bar\theta) = \left(-\frac{\partial}{\partial\theta_\alpha} - i\bar\theta_{\dot\beta}\bar\sigma^{\mu\dot\beta\alpha}\partial_\mu\right)\left(\frac{\partial}{\partial\theta^\alpha} + i\sigma^\nu{}_{\alpha\dot\gamma}\bar\theta^{\dot\gamma}\partial_\nu\right)\times$$

$$\times\left[\sqrt{2}\theta^\rho\psi_\rho(x) - \frac{i}{\sqrt{2}}\theta\theta\partial_\eta\psi(x)\sigma^\eta\bar\theta\right] =$$

$$= \left(-\frac{\partial}{\partial \theta_\alpha} - i\bar\theta_{\dot\beta}\bar\sigma^{\mu\dot\beta\alpha}\partial_\mu\right)\left[\sqrt{2}\psi_\alpha(x) - \frac{i}{\sqrt{2}}2\theta_\alpha\partial_\eta\psi(x)\sigma^\eta\bar\theta + i\sqrt{2}\sigma^\nu{}_{\alpha\dot\gamma}\bar\theta^{\dot\gamma}\theta^\rho\partial_\nu\psi_\rho(x) + \right.$$
$$\left. +\frac{1}{\sqrt{2}}\theta\theta\sigma^\nu{}_{\alpha\dot\gamma}\bar\theta^{\dot\gamma}\partial_\nu\partial_\eta\psi(x)\sigma^\eta\bar\theta\right] =$$
$$= \frac{i}{\sqrt{2}}4\partial_\eta\psi\sigma^\eta\bar\theta - i\sqrt{2}\sigma^\nu{}_{\alpha\dot\gamma}\bar\theta^{\dot\gamma}\partial_\nu\psi(x) + \frac{1}{\sqrt{2}}2\theta^\alpha\sigma^\nu{}_{\alpha\dot\gamma}\bar\theta^{\dot\gamma}\partial_\nu\partial_\eta\psi(x)\sigma^\eta\bar\theta -$$
$$-i\sqrt{2}\bar\theta_{\dot\beta}\bar\sigma^{\mu\dot\beta\alpha}\partial_\mu\psi_\alpha(x) - \sqrt{2}\bar\theta_{\dot\beta}\bar\sigma^{\mu\dot\beta\alpha}\theta_\alpha\partial_\mu\partial_\eta\psi(x)\sigma^\eta\bar\theta + \sqrt{2}\bar\theta_{\dot\beta}\bar\sigma^{\mu\dot\beta\alpha}\sigma^\nu{}_{\alpha\dot\gamma}\bar\theta^{\dot\gamma}\theta^\rho\partial_\mu\partial_\nu\psi_\rho(x)$$
$$-\frac{i}{\sqrt{2}}\bar\theta_{\dot\beta}\bar\sigma^{\mu\dot\beta\alpha}\sigma^\nu{}_{\alpha\dot\gamma}\bar\theta^{\dot\gamma}\theta\theta\partial_\mu\partial_\mu\partial_\eta\psi(x)\sigma^\eta\bar\theta. \quad (2.19)$$

Taking into account that all terms containing the third degrees of the variables $\theta$, $\bar\theta$ or higher are zero [16], we simplify (2.19):

$$D^\alpha D_\alpha \Phi^{(\psi)}(x,\theta,\bar\theta) =$$
$$= i2\sqrt{2}\partial_\eta\psi\sigma^\eta\bar\theta - i\sqrt{2}\left(\sigma^\nu\bar\theta\right)_\alpha\partial_\nu\psi^\alpha(x) + \sqrt{2}\left(\theta\sigma^\nu\bar\theta\right)\partial_\nu\partial_\eta\psi(x)\sigma^\eta\bar\theta -$$
$$-i\sqrt{2}\left(\bar\theta\bar\sigma\right)^\alpha\partial_\mu\psi_\alpha(x) - \sqrt{2}\left(\bar\theta\bar\sigma^\mu\theta\right)\partial_\mu\partial_\eta\psi(x)\sigma^\eta\bar\theta + \sqrt{2}\left(\bar\theta\bar\sigma^\mu\right)^\alpha\left(\sigma^\nu\bar\theta\right)_\alpha\theta^\rho\partial_\mu\partial_\nu\psi_\rho(x).$$
$$(2.20)$$

Further, we use the following identity

$$\bar\theta_{\dot\beta}\bar\sigma^{\mu\dot\beta\alpha}\sigma^\nu{}_{\alpha\dot\gamma}\bar\theta^{\dot\gamma}\partial_\mu\partial_\nu = \bar\theta_{\dot\beta}\left(\frac{1}{2}\left\{\bar\sigma^{\mu\dot\beta\alpha},\sigma^\nu{}_{\alpha\dot\gamma}\right\} + \frac{1}{2}\left[\bar\sigma^{\mu\dot\beta\alpha},\sigma^\nu{}_{\alpha\dot\gamma}\right]\right)\bar\theta^{\dot\gamma}\partial_\mu\partial_\nu = $$
$$= -\frac{1}{2}2g^{\mu\nu}\delta^{\dot\beta}_{\dot\gamma}\bar\theta_{\dot\beta}\bar\theta^{\dot\gamma}\partial_\mu\partial_\nu = -\bar\theta\bar\theta\Box. \quad (2.21)$$

Then, taking into account the identity (2.21) and collecting similar terms, we finally get

$$D^\alpha D_\alpha \Phi^{(\psi)}(x,\theta,\bar\theta) =$$
$$= i2\sqrt{2}\partial_\eta\psi\sigma^\eta\bar\theta + i2\sqrt{2}\partial_\nu\psi(x)\sigma^\nu\bar\theta + \sqrt{2}\left(\theta\sigma^\nu\bar\theta\right)\partial_\nu\partial_\eta\psi(x)\sigma^\eta\bar\theta +$$
$$+\sqrt{2}\left(\theta\sigma^\mu\bar\theta\right)\partial_\mu\partial_\eta\psi(x)\sigma^\eta\bar\theta - \sqrt{2}\theta\theta\Box\psi(x) =$$
$$= i4\sqrt{2}\partial_\eta\psi\sigma^\eta\bar\theta + 2\sqrt{2}\left(\theta\sigma^\nu\bar\theta\right)\partial_\nu\partial_\eta\psi(x)\sigma^\eta\bar\theta - \sqrt{2}\theta\theta\Box\psi(x). \quad (2.22)$$

Similarly, we find the expressions for $D^\alpha D_\alpha \Phi^{(A)}(x,\theta,\bar\theta)$ and $D^\alpha D_\alpha \Phi^{(F)}(x,\theta,\bar\theta)$. So, for example, in the first case we find

$$D^{\alpha}D_{\alpha}\Phi^{(A)}(x,\theta,\bar{\theta}) = \left(-\frac{\partial}{\partial\theta_{\alpha}} - i\bar{\theta}_{\dot{\beta}}\bar{\sigma}^{\mu\dot{\beta}\alpha}\partial_{\mu}\right)\left(\frac{\partial}{\partial\theta^{\alpha}} + i\sigma^{\nu}{}_{\alpha\dot{\gamma}}\bar{\theta}^{\dot{\gamma}}\partial_{\nu}\right) \times$$

$$\times\left[A(x) + i\theta^{\lambda}\sigma^{\eta}{}_{\lambda\dot{\lambda}}\bar{\theta}^{\dot{\lambda}}\partial_{\eta}A(x) + \frac{1}{4}\theta\theta\bar{\theta}\bar{\theta}\Box A(x)\right] =$$

$$= \left(-\frac{\partial}{\partial\theta_{\alpha}} - i\bar{\theta}_{\dot{\beta}}\bar{\sigma}^{\mu\dot{\beta}\alpha}\partial_{\mu}\right)\left[i\sigma^{\eta}{}_{\alpha\dot{\lambda}}\bar{\theta}^{\dot{\lambda}}\partial_{\eta}A(x) + \frac{1}{4}2\theta_{\alpha}\bar{\theta}\bar{\theta}\Box A(x) + i\sigma^{\nu}{}_{\alpha\dot{\gamma}}\bar{\theta}^{\dot{\gamma}}\partial_{\nu}A(x) -\right.$$

$$\left. -\sigma^{\nu}{}_{\alpha\dot{\gamma}}\bar{\theta}^{\dot{\gamma}}\theta^{\lambda}\sigma^{\eta}{}_{\lambda\dot{\lambda}}\bar{\theta}^{\dot{\lambda}}\partial_{\nu}\partial_{\eta}A(x)\right] =$$

$$= -\frac{1}{2}2\bar{\theta}\bar{\theta}\Box A(x) + \sigma^{\nu}{}_{\alpha\dot{\gamma}}\bar{\theta}^{\dot{\gamma}}\varepsilon^{\alpha\lambda}\sigma^{\eta}{}_{\lambda\dot{\lambda}}\bar{\theta}^{\dot{\lambda}}\partial_{\nu}\partial_{\eta}A(x) + \bar{\theta}_{\dot{\beta}}\bar{\sigma}^{\mu\dot{\beta}\alpha}\sigma^{\eta}{}_{\alpha\dot{\lambda}}\bar{\theta}^{\dot{\lambda}}\partial_{\mu}\partial_{\eta}A(x) +$$

$$+\bar{\theta}_{\dot{\beta}}\bar{\sigma}^{\mu\dot{\beta}\alpha}\sigma^{\nu}{}_{\alpha\dot{\gamma}}\bar{\theta}^{\dot{\gamma}}\partial_{\mu}\partial_{\nu}A(x) =$$

$$= -\bar{\theta}\bar{\theta}\Box A(x) + \left(\sigma^{\nu}\bar{\theta}\right)_{\alpha}\varepsilon^{\alpha\lambda}\left(\sigma^{\eta}\bar{\theta}\right)_{\lambda}\partial_{\nu}\partial_{\eta}A(x) + 2\left(\bar{\theta}\bar{\sigma}^{\mu}\right)^{\alpha}\left(\sigma^{\eta}\bar{\theta}\right)_{\alpha}\partial_{\mu}\partial_{\eta}A(x). \quad (2.23)$$

To simplify this expression, we use the following useful relations

$$\bar{\sigma}^{\nu\dot{\alpha}\alpha} = \varepsilon^{\dot{\alpha}\dot{\beta}}\varepsilon^{\alpha\beta}\sigma^{\nu}{}_{\beta\dot{\beta}},$$

$$-2g^{\nu\mu} = \bar{\sigma}^{\nu\dot{\alpha}\alpha}\sigma^{\mu}{}_{\alpha\dot{\alpha}} = \varepsilon^{\dot{\alpha}\dot{\beta}}\varepsilon^{\alpha\beta}\sigma^{\nu}{}_{\beta\dot{\beta}}\sigma^{\mu}{}_{\alpha\dot{\alpha}} = -\varepsilon^{\dot{\alpha}\dot{\beta}}\sigma^{\nu}{}_{\beta\dot{\beta}}\varepsilon^{\beta\alpha}\sigma^{\mu}{}_{\alpha\dot{\alpha}},$$

$$2\varepsilon_{\dot{\beta}\dot{\alpha}}g^{\nu\mu} = \varepsilon_{\dot{\beta}\dot{\alpha}}\varepsilon^{\dot{\alpha}\dot{\beta}}\sigma^{\nu}{}_{\beta\dot{\beta}}\varepsilon^{\beta\alpha}\sigma^{\mu}{}_{\alpha\dot{\alpha}} = 2\sigma^{\nu}{}_{\beta\dot{\beta}}\varepsilon^{\beta\alpha}\sigma^{\mu}{}_{\alpha\dot{\alpha}}, \quad (2.24)$$

$$\sigma^{\nu}{}_{\beta\dot{\beta}}\varepsilon^{\beta\alpha}\sigma^{\mu}{}_{\alpha\dot{\alpha}} = \varepsilon_{\dot{\beta}\dot{\alpha}}g^{\nu\mu}.$$

Taking into account the formulae (2.24), (2.21) and collecting similar terms, we obtain the explicit expressions for $D^{\alpha}D_{\alpha}\Phi^{(A)}(x,\theta,\bar{\theta})$ and $D^{\alpha}D_{\alpha}\Phi^{(F)}(x,\theta,\bar{\theta})$, correspondingly

$$D^{\alpha}D_{\alpha}\Phi^{(A)}(x,\theta,\bar{\theta}) = -\bar{\theta}\bar{\theta}\Box A(x) + \left(\sigma^{\nu}\bar{\theta}\right)_{\alpha}\varepsilon^{\alpha\lambda}\left(\sigma^{\eta}\bar{\theta}\right)_{\lambda}\partial_{\nu}\partial_{\eta}A(x) +$$

$$+2\left(\bar{\theta}\bar{\sigma}^{\mu}\right)^{\alpha}\left(\sigma^{\eta}\bar{\theta}\right)_{\alpha}\partial_{\mu}\partial_{\eta}A(x) =$$

$$= -\bar{\theta}\bar{\theta}\Box A(x) + \varepsilon_{\dot{\gamma}\dot{\lambda}}\bar{\theta}^{\dot{\gamma}}\bar{\theta}^{\dot{\lambda}}g^{\nu\eta}\partial_{\nu}\partial_{\eta}A(x) - 2\bar{\theta}\bar{\theta}\Box A(x) = \quad (2.25)$$

$$= -\bar{\theta}\bar{\theta}\Box A(x) - \bar{\theta}\bar{\theta}\Box A(x) - 2\bar{\theta}\bar{\theta}\Box A(x) =$$

$$= -4\bar{\theta}\bar{\theta}\Box A(x);$$

$$D^{\alpha}D_{\alpha}\Phi^{(F)}(x,\theta,\bar{\theta}) = \left(-\frac{\partial}{\partial\theta_{\alpha}} - i\bar{\theta}_{\dot{\beta}}\bar{\sigma}^{\mu\dot{\beta}\alpha}\partial_{\mu}\right)\left(\frac{\partial}{\partial\theta^{\alpha}} + i\sigma^{\nu}{}_{\alpha\dot{\gamma}}\bar{\theta}^{\dot{\gamma}}\partial_{\nu}\right)[\theta\theta F(x)] =$$

$$= \left(-\frac{\partial}{\partial \theta_\alpha} - i\bar{\theta}_{\dot\beta}\bar{\sigma}^{\mu\dot\beta\alpha}\partial_\mu\right)\left[2\theta_\alpha F(x) + i\sigma^\nu{}_{\alpha\dot\gamma}\bar{\theta}^{\dot\gamma}\theta\theta\partial_\nu F(x)\right] =$$

$$= -4F(x) + 2i\theta^\alpha\sigma^\nu{}_{\alpha\dot\gamma}\bar{\theta}^{\dot\gamma}\partial_\nu F(x) - 2i\bar{\theta}_{\dot\beta}\bar{\sigma}^{\mu\dot\beta\alpha}\theta_\alpha\partial_\mu F(x) + \bar{\theta}_{\dot\beta}\bar{\sigma}^{\mu\dot\beta\alpha}\sigma^\nu{}_{\alpha\dot\gamma}\bar{\theta}^{\dot\gamma}\theta\theta\partial_\mu\partial_\nu F(x) =$$

$$= -4F(x) + 4i(\theta\sigma^\nu\bar{\theta})\partial_\nu F(x) + (\bar{\theta}\bar{\sigma}^\mu)^\alpha(\sigma^\nu\bar{\theta})_\alpha \theta\theta\partial_\mu\partial_\nu F(x) =$$

$$= -4F(x) + 4i(\theta\sigma^\nu\bar{\theta})\partial_\nu F(x) + \bar{\theta}\bar{\theta}\theta\theta\Box F(x). \qquad (2.26)$$

## 2.4 The interaction Lagrangian of a superparticle with an external chiral matter superfield

Now back to the superfield $S(x,\theta,\bar{\theta})$. As mentioned above, it actually plays the role of the prepotential of the chiral superfield $\Phi(x,\theta,\bar{\theta})$. Let us write again the connection between the superfields $S(x,\theta,\bar{\theta})$ and $\Phi(x,\theta,\bar{\theta})$:

$$S(x,\theta,\bar{\theta}) = \frac{1}{16\Box}D^2\Phi(x,\theta,\bar{\theta}). \qquad (2.27)$$

Making use of the integral representation of the D'Alembert operator in terms of the Green's function, we rewrite expression (2.27) as follows

$$S(x,\theta,\bar{\theta}) = \frac{1}{16}\int G(x,y)D^2\Phi(y,\theta,\bar{\theta})dy. \qquad (2.28)$$

Next, using the given representation we find

$$\bar{\sigma}^{\mu\dot\beta\alpha}\bar{D}_{\dot\beta}D_\alpha S(x,\theta,\bar{\theta}) = \frac{1}{16}\bar{\sigma}^{\mu\dot\beta\alpha}\bar{D}_{\dot\beta}D_\alpha\int G(x,y)D^2\Phi(y,\theta,\bar{\theta})dy. \qquad (2.29)$$

The expression on the left-hand side of (2.29) is included in the initial interaction Lagrangian (1.29). Finally, using the explicit expression obtained above, for example, for $D^\alpha D_\alpha\Phi^{(\psi)}(x,\theta,\bar{\theta})$, Eq. (2.22), we get

$$\bar{\sigma}^{\mu\dot\beta\alpha}\bar{D}_{\dot\beta}D_\alpha S^{(\psi)}(x,\theta,\bar{\theta}) = \frac{1}{16}\bar{\sigma}^{\mu\dot\beta\alpha}\left(-\frac{\partial}{\partial\bar{\theta}^{\dot\beta}} - i\theta^\beta\sigma^\lambda{}_{\beta\dot\beta}\partial_\lambda\right)\left(\frac{\partial}{\partial\theta^\alpha} + i\sigma^\rho{}_{\alpha\dot\alpha}\bar{\theta}^{\dot\alpha}\partial_\rho\right)\times$$

$$\times \int G(x,y)\Phi^{(\psi)}(y,\theta,\bar{\theta})dy = \frac{1}{16}\bar{\sigma}^{\mu\dot\beta\alpha}\left(-\frac{\partial}{\partial\bar{\theta}^{\dot\beta}} - i\theta^\beta\sigma^\lambda{}_{\beta\dot\beta}\partial_\lambda\right)\left(\frac{\partial}{\partial\theta^\alpha} + i\sigma^\rho{}_{\alpha\dot\alpha}\bar{\theta}^{\dot\alpha}\partial_\rho\right)\times$$

$$\times\left[\int G(x,y)\{i4\sqrt{2}\partial_\eta\psi(y)\sigma^\eta\bar{\theta} + 2\sqrt{2}\theta^\gamma\sigma^\nu{}_{\gamma\dot\gamma}\bar{\theta}^{\dot\gamma}\partial_\nu\partial_\eta\psi(y)\sigma^\eta\bar{\theta}\}dy - \sqrt{2}\theta\theta\theta\psi(x)\right] =$$

$$= \frac{1}{16} \bar{\sigma}^{\mu\dot{\beta}\alpha} \left(-\frac{\partial}{\partial \bar{\theta}^{\dot{\beta}}} - i\theta^{\beta} \sigma^{\lambda}{}_{\beta\dot{\beta}} \partial_{\lambda}\right) \left[2\sqrt{2} \int G(x,y) \sigma^{\nu}{}_{\alpha\dot{\gamma}} \bar{\theta}^{\dot{\gamma}} \partial_{\nu}\partial_{\eta}\psi(y)\sigma^{\eta}\bar{\theta} dy - \sqrt{2\theta\theta}\psi_{\alpha}(x)\right.$$

$$\left. -4\sqrt{2} \int \partial_{\rho} G(x,y) \sigma^{\rho}{}_{\alpha\dot{\alpha}} \bar{\theta}^{\dot{\alpha}} \partial_{\eta}\psi(y)\sigma^{\eta}\bar{\theta} dy\right] =$$

$$= \frac{1}{16}\left[-2\sqrt{2} \int G(x,y) \bar{\sigma}^{\mu\dot{\beta}\alpha}\{\sigma^{\nu}{}_{\alpha\dot{\beta}} \partial_{\nu}\partial_{\eta}\psi(y)\sigma^{\eta}\bar{\theta} + \sigma^{\nu}{}_{\alpha\dot{\gamma}}\bar{\theta}^{\dot{\gamma}}\partial_{\nu}\partial_{\eta}(\psi(y)\sigma^{\eta})_{\dot{\beta}}\}dy - \right.$$

$$-2\sqrt{2}\bar{\theta}_{\dot{\beta}}\bar{\sigma}^{\mu\dot{\beta}\alpha}\psi_{\alpha}(x) +$$

$$+4\sqrt{2}\int \partial_{\rho}G(x,y)\bar{\sigma}^{\mu\dot{\beta}\alpha}\{\sigma^{\rho}{}_{\alpha\dot{\beta}}\partial_{\eta}\psi(y)\sigma^{\eta}\bar{\theta} + \sigma^{\rho}{}_{\alpha\dot{\alpha}}\bar{\theta}^{\dot{\alpha}}\partial_{\eta}(\psi(y)\sigma^{\eta})_{\dot{\beta}}\}dy -$$

$$-i2\sqrt{2}\int \partial_{\lambda}G(x,y)\theta^{\beta}\sigma^{\lambda}{}_{\beta\dot{\beta}}\bar{\sigma}^{\mu\dot{\beta}\alpha}\sigma^{\nu}{}_{\alpha\dot{\gamma}}\bar{\theta}^{\dot{\gamma}}\partial_{\nu}\partial_{\eta}\psi(y)\sigma^{\eta}\bar{\theta}dy + i\sqrt{2}\theta^{\beta}\sigma^{\lambda}{}_{\beta\dot{\beta}}\bar{\sigma}^{\mu\dot{\beta}\alpha}\bar{\theta}\bar{\theta}\partial_{\lambda}\psi_{\alpha}(x)$$

$$\left. + i4\sqrt{2}\int \partial_{\lambda}\partial_{\rho}G(x,y)\theta^{\beta}\sigma^{\lambda}{}_{\beta\dot{\beta}}\bar{\sigma}^{\mu\dot{\beta}\alpha}\sigma^{\rho}{}_{\alpha\dot{\alpha}}\bar{\theta}^{\dot{\alpha}}\partial_{\eta}\psi(y)\sigma^{\eta}\bar{\theta}dy\right]. \quad (2.30)$$

Collecting similar terms in (2.30), we find the following somewhat cumbersome expression

$$\bar{\sigma}^{\mu\dot{\beta}\alpha}\bar{D}_{\dot{\beta}}D_{\alpha}S^{(\psi)}(x,\theta,\bar{\theta}) = \frac{1}{16}\left[4\sqrt{2}\int G(x,y)\partial^{\mu}\partial_{\eta}\psi(y)\sigma^{\eta}\bar{\theta}dy - \right.$$

$$-2\sqrt{2}\int G(x,y)\sigma^{\eta}{}_{\beta\dot{\beta}}\bar{\sigma}^{\mu\dot{\beta}\alpha}\sigma^{\nu}{}_{\alpha\dot{\gamma}}\bar{\theta}^{\dot{\gamma}}\partial_{\nu}\partial_{\eta}\psi^{\beta}(y)dy - 2\sqrt{2}\bar{\theta}_{\dot{\beta}}\bar{\sigma}^{\mu\dot{\beta}\alpha}\psi_{\alpha}(x) -$$

$$-8\sqrt{2}\int \partial^{\mu}G(x,y)\partial_{\eta}\psi(y)\sigma^{\eta}\bar{\theta}dy + 4\sqrt{2}\int \partial_{\rho}G(x,y)\sigma^{\eta}{}_{\beta\dot{\beta}}\bar{\sigma}^{\mu\dot{\beta}\alpha}\sigma^{\rho}{}_{\alpha\dot{\alpha}}\bar{\theta}^{\dot{\alpha}}\partial_{\eta}\psi^{\beta}(y)dy -$$

$$-i2\sqrt{2}\int \partial_{\lambda}G(x,y)\theta^{\beta}\sigma^{\lambda}{}_{\beta\dot{\beta}}\bar{\sigma}^{\mu\dot{\beta}\alpha}\sigma^{\nu}{}_{\alpha\dot{\gamma}}\bar{\theta}^{\dot{\gamma}}\partial_{\nu}\partial_{\eta}\psi(y)\sigma^{\eta}\bar{\theta}dy + i\sqrt{2}\theta^{\beta}\sigma^{\lambda}{}_{\beta\dot{\beta}}\bar{\sigma}^{\mu\dot{\beta}\alpha}\bar{\theta}\bar{\theta}\partial_{\lambda}\psi_{\alpha}(x) +$$

$$\left. +i4\sqrt{2}\int \partial_{\lambda}\partial_{\rho}G(x,y)\theta^{\beta}\sigma^{\lambda}{}_{\beta\dot{\beta}}\bar{\sigma}^{\mu\dot{\beta}\alpha}\sigma^{\rho}{}_{\alpha\dot{\alpha}}\bar{\theta}^{\dot{\alpha}}\partial_{\eta}\psi(y)\sigma^{\eta}\bar{\theta}dy\right].$$

(2.31)

Here we need some additional spinor relations. Combining and transforming formulae (A.10), (A.11) (see Appendix), as well as (A.8) and (A.9), we obtain the following useful expressions for the $\sigma$-matrices:

$$\sigma^{\lambda}\bar{\sigma}^{\mu}\sigma^{\nu} = \left(g^{\lambda\nu}\sigma^{\mu} - g^{\mu\nu}\sigma^{\lambda} - g^{\lambda\mu}\sigma^{\nu} + i\varepsilon^{\lambda\mu\nu\rho}\sigma_{\rho}\right), \quad (2.32)$$

$$\sigma^{\lambda}{}_{\beta\dot{\beta}}\bar{\sigma}^{\mu\dot{\beta}\alpha} = -g^{\lambda\mu}\delta^{\alpha}_{\beta} + 2\sigma^{\lambda\mu\alpha}{}_{\beta}. \quad (2.33)$$

We consider each of the eight terms in parentheses in the expression (2.31) separately, to simplify them as much as possible. We leave the first term unchanged:

$$4\sqrt{2}\int G(x,y)\,\partial^{\mu}\partial_{\eta}(\psi(y)\sigma^{\eta}\bar{\theta})dy. \quad (2.31a)$$

We transform the second term making use of formula (2.32):

$$-2\sqrt{2}\int G(x,y)\sigma^{\eta}{}_{\beta\dot{\beta}}\bar{\sigma}^{\mu\dot{\beta}\alpha}\sigma^{\nu}{}_{\alpha\dot{\gamma}}\bar{\theta}^{\dot{\gamma}}\partial_{\nu}\partial_{\eta}\psi^{\beta}(y)dy=$$
$$=2\sqrt{2}\int G(x,y)\partial_{\nu}\partial_{\eta}\psi^{\beta}(y)\left[g^{\eta\nu}\sigma^{\mu}{}_{\beta\dot{\gamma}}-g^{\mu\nu}\sigma^{\eta}{}_{\beta\dot{\gamma}}-g^{\eta\mu}\sigma^{\nu}{}_{\beta\dot{\gamma}}+i\varepsilon^{\eta\mu\nu\lambda}\sigma_{\lambda\beta\dot{\gamma}}\right]\bar{\theta}^{\dot{\gamma}}dy=$$
$$=2\sqrt{2}\psi(x)\sigma^{\mu}\bar{\theta}-2\sqrt{2}\int G(x,y)\partial^{\mu}\partial_{\eta}\left(\psi(y)\sigma^{\eta}\bar{\theta}\right)dy-2\sqrt{2}\int G(x,y)\partial_{\nu}\partial^{\mu}\left(\psi(y)\sigma^{\nu}\bar{\theta}\right)dy.$$
(2.31b)

The third and fourth terms are left unchanged:

$$-2\sqrt{2}\theta\bar{\sigma}^{\mu}\psi(x);\tag{2.31c}$$

$$-8\sqrt{2}\int\partial^{\mu}G(x,y)\partial_{\eta}\left(\psi(y)\sigma^{\eta}\bar{\theta}\right)dy.\tag{2.31d}$$

Further we transform the fifth term by using the formula (2.32) again:

$$4\sqrt{2}\int\partial_{\rho}G(x,y)\sigma^{\eta}{}_{\beta\dot{\beta}}\bar{\sigma}^{\mu\dot{\beta}\alpha}\sigma^{\rho}{}_{\alpha\dot{\alpha}}\bar{\theta}^{\dot{\alpha}}\partial_{\eta}\psi^{\beta}(y)dy=$$
$$=-4\sqrt{2}\int\partial_{\rho}G(x,y)\partial_{\eta}\psi^{\beta}(y)\left[g^{\eta\rho}\sigma^{\mu}{}_{\beta\dot{\alpha}}-g^{\mu\rho}\sigma^{\eta}{}_{\beta\dot{\alpha}}-g^{\eta\mu}\sigma^{\rho}{}_{\beta\dot{\alpha}}+i\varepsilon^{\eta\mu\rho\nu}\sigma_{\nu\beta\dot{\alpha}}\right]\bar{\theta}^{\dot{\alpha}}dy=$$
$$-4\sqrt{2}\int\partial_{\rho}G(x,y)\partial^{\rho}\left(\psi(y)\sigma^{\mu}\bar{\theta}\right)dy+4\sqrt{2}\int\partial^{\mu}G(x,y)\partial_{\eta}\left(\psi(y)\sigma^{\eta}\bar{\theta}\right)dy+\quad(2.31e)$$
$$+4\sqrt{2}\int\partial_{\rho}G(x,y)\partial^{\mu}\left(\psi(y)\sigma^{\rho}\bar{\theta}\right)dy-i4\sqrt{2}\int\varepsilon^{\eta\mu\rho\nu}\partial_{\rho}G(x,y)\partial_{\eta}\left(\psi(y)\sigma_{\nu}\bar{\theta}\right)dy.$$

The sixth term is transformed similarly to the fifth one

$$-i2\sqrt{2}\int\partial_{\lambda}G(x,y)\theta^{\beta}\sigma^{\lambda}{}_{\beta\dot{\beta}}\bar{\sigma}^{\mu\dot{\beta}\alpha}\sigma^{\nu}{}_{\alpha\dot{\gamma}}\bar{\theta}^{\dot{\gamma}}\partial_{\nu}\partial_{\eta}\psi(y)\sigma^{\eta}\bar{\theta}dy=$$
$$=-i2\sqrt{2}\int\partial_{\lambda}G(x,y)\theta^{\beta}\left[g^{\lambda\nu}\sigma^{\mu}{}_{\beta\dot{\gamma}}-g^{\mu\nu}\sigma^{\lambda}{}_{\beta\dot{\gamma}}-g^{\lambda\mu}\sigma^{\nu}{}_{\beta\dot{\gamma}}+i\varepsilon^{\lambda\mu\nu\rho}\sigma_{\rho\beta\dot{\gamma}}\right]\bar{\theta}^{\dot{\gamma}}\partial_{\nu}\partial_{\eta}\psi(y)\sigma^{\eta}\bar{\theta}dy=$$
$$=-i2\sqrt{2}\int\partial^{\nu}G(x,y)\left(\theta\sigma^{\mu}\bar{\theta}\right)\partial_{\nu}\partial_{\eta}\psi(y)\sigma^{\eta}\bar{\theta}dy+i2\sqrt{2}\int\partial_{\lambda}G(x,y)\left(\theta\sigma^{\lambda}\bar{\theta}\right)\partial^{\mu}\partial_{\eta}\psi(y)\sigma^{\eta}\bar{\theta}dy+$$
$$+i2\sqrt{2}\int\partial^{\mu}G(x,y)\left(\theta\sigma^{\nu}\bar{\theta}\right)\partial_{\nu}\partial_{\eta}\psi(y)\sigma^{\eta}\bar{\theta}dy+2\sqrt{2}\int\varepsilon^{\lambda\mu\nu\rho}\partial_{\lambda}G(x,y)\left(\theta\sigma_{\rho}\bar{\theta}\right)\partial_{\nu}\partial_{\eta}\psi(y)\sigma^{\eta}\bar{\theta}dy.$$
(2.31f)

To transform the seventh term, we need to use the formula (2.33):

$$i\sqrt{2}\theta^{\beta}\sigma^{\lambda}{}_{\beta\dot{\beta}}\bar{\sigma}^{\mu\dot{\beta}\alpha}\bar{\theta}\bar{\theta}\partial_{\lambda}\psi_{\alpha}(x)=i\sqrt{2}\theta^{\beta}\left(-g^{\lambda\mu}\delta^{\alpha}_{\beta}+2\sigma^{\lambda\mu\alpha}{}_{\beta}\right)\bar{\theta}\bar{\theta}\partial_{\lambda}\psi_{\alpha}(x)=$$
$$=-i\sqrt{2}\bar{\theta}\bar{\theta}\partial^{\mu}\left(\theta\psi(x)\right)+i2\sqrt{2}\bar{\theta}\bar{\theta}\partial_{\lambda}\left(\theta\sigma^{\lambda\mu}\psi(x)\right).\tag{2.31g}$$

Finally, the eighth term is transformed by the use of the formula (2.32):

$$i4\sqrt{2}\int \partial_\lambda \partial_\rho G(x,y)\theta^\beta \sigma^\lambda{}_{\beta\dot\beta}\bar\sigma^{\mu\dot\beta\alpha}\sigma^\rho{}_{\alpha\dot\alpha}\bar\theta^{\dot\alpha}\partial_\eta\psi(y)\sigma^\eta\bar\theta dy =$$

$$= i4\sqrt{2}\int \partial_\lambda \partial_\rho G(x,y)\theta^\beta \left(g^{\lambda\rho}\sigma^\mu{}_{\beta\dot\alpha} - g^{\mu\rho}\sigma^\lambda{}_{\beta\dot\alpha} - g^{\lambda\mu}\sigma^\rho{}_{\beta\dot\alpha} + i\varepsilon^{\lambda\mu\rho\nu}\sigma_{\nu\beta\dot\alpha}\right)\bar\theta^{\dot\alpha}\partial_\eta\psi(y)\sigma^\eta\bar\theta dy =$$

$$= i4\sqrt{2}\left(\theta\sigma^\mu\bar\theta\right)\partial_\eta\psi(x)\sigma^\eta\bar\theta - i4\sqrt{2}\int \partial_\lambda \partial^\mu G(x,y)\left(\theta\sigma^\lambda\bar\theta\right)\partial_\eta\psi(y)\sigma^\eta\bar\theta dy -$$

$$-i4\sqrt{2}\int \partial^\mu \partial_\rho G(x,y)\left(\theta\sigma^\rho\bar\theta\right)\partial_\eta\psi(y)\sigma^\eta\bar\theta dy.$$

(2.31h)

Gathering the expressions (2.31a)- (2.31h) we end up with

$$\bar\sigma^{\mu\dot\beta\alpha}\bar D_{\dot\beta}D_\alpha S^{(\psi)}(x,\theta,\bar\theta) = \frac{1}{16}\Bigl[4\sqrt{2}\psi(x)\sigma^\mu\bar\theta - i\sqrt{2}\theta\bar\theta\partial^\mu\left(\theta\psi(x)\right) +$$

$$+i2\sqrt{2}\theta\bar\theta\partial_\nu\left(\theta\sigma^{\nu\mu}\psi(x)\right) + i4\sqrt{2}\left(\theta\sigma^\mu\bar\theta\right)\partial_\nu\psi(x)\sigma^\nu\bar\theta -$$

$$-4\sqrt{2}\Bigl\{\int \partial_\nu G(x,y)\partial^\nu\left(\psi(y)\sigma^\mu\bar\theta\right)dy + \int \partial^\mu G(x,y)\partial_\nu\left(\psi(y)\sigma^\nu\bar\theta\right)dy -$$

$$-\int \partial_\nu G(x,y)\partial^\mu\left(\psi(y)\sigma^\nu\bar\theta\right)dy + i\int \varepsilon^{\nu\mu\lambda\rho}\partial_\lambda G(x,y)\partial_\nu\left(\psi(y)\sigma_\rho\bar\theta\right)dy\Bigr\} +$$

$$+i2\sqrt{2}\left(\theta\sigma^\nu\bar\theta\right)\Bigl\{\int \partial_\nu G(x,y)\partial^\mu\partial_\eta\psi(y)\sigma^\eta\bar\theta dy - \int g_{\mu\nu}\partial^\nu G(x,y)\partial_\nu\partial_\eta\psi(y)\sigma^\eta\bar\theta dy +$$

$$+\int \partial^\mu G(x,y)\partial_\nu\partial_\eta\psi(y)\sigma^\eta\bar\theta dy - i\int \varepsilon^{\rho\mu\lambda}{}_\nu\partial_\rho G(x,y)\partial_\lambda\partial_\eta\psi(y)\sigma^\eta\bar\theta dy -$$

$$-4\int \partial^\mu\partial_\nu G(x,y)\partial_\eta\psi(y)\sigma^\eta\bar\theta dy\Bigr\}\Bigr].$$

(2.34)

Taking into consideration (2.25) it is easy to make sure that a similar expression for $S^{(A)}(x,\theta,\bar\theta)$ equals zero:

$$\bar\sigma^{\mu\dot\beta\alpha}\bar D_{\dot\beta}D_\alpha S^{(A)}(x,\theta,\bar\theta) = \frac{1}{16}\bar\sigma^{\mu\dot\beta\alpha}\left(-\frac{\partial}{\partial\bar\theta^{\dot\beta}} - i\theta^\beta\sigma^\lambda{}_{\beta\dot\beta}\partial_\lambda\right)\left(\frac{\partial}{\partial\theta^\alpha} + i\sigma^\rho{}_{\alpha\dot\alpha}\bar\theta^{\dot\alpha}\partial_\rho\right)\times$$

$$\times \int G(x,y)\Phi^{(A)}(y,\theta,\bar\theta)dy =$$

$$= \frac{1}{16}\bar\sigma^{\mu\dot\beta\alpha}\left(-\frac{\partial}{\partial\bar\theta^{\dot\beta}} - i\theta^\beta\sigma^\lambda{}_{\beta\dot\beta}\partial_\lambda\right)\left(\frac{\partial}{\partial\theta^\alpha} + i\sigma^\rho{}_{\alpha\dot\alpha}\bar\theta^{\dot\alpha}\partial_\rho\right)\left[-4\theta\bar\theta A(x)\right] = 0.$$

(2.35)

Let us calculate the second term of the interaction Lagrangian (1.29), namely

$$-i\dot\theta^\alpha D_\alpha S(x,\theta,\bar\theta) = -\frac{i}{16}\dot\theta^\alpha D_\alpha \int G(x,y)D^2\Phi(y,\theta,\bar\theta)dy.$$ (2.36)

For the contributions with the fields $\psi(x)$ and $A(x)$ we have an explicit form for covariant derivative:

$$-i\dot{\theta}^{\alpha} D_{\alpha} S^{(\psi)}(x,\theta,\bar{\theta}) = -\frac{i}{16}\dot{\theta}^{\alpha} D_{\alpha} \int G(x,y) D^{2}\Phi^{(\psi)}(y,\theta,\bar{\theta}) dy =$$

$$= -\frac{i}{16}\dot{\theta}^{\alpha}\left(\frac{\partial}{\partial\theta^{\alpha}} + i\sigma^{\mu}{}_{\alpha\dot{\beta}}\bar{\theta}^{\dot{\beta}}\partial_{\mu}\right) \times$$

$$\times\left[\int G(x,y)\left\{i4\sqrt{2}\partial_{\eta}\psi(y)\sigma^{\eta}\bar{\theta} + 2\sqrt{2}\theta^{\beta}\sigma^{\nu}{}_{\beta\dot{\alpha}}\bar{\theta}^{\dot{\alpha}}\partial_{\nu}\partial_{\eta}\psi(y)\sigma^{\eta}\bar{\theta}\right\} dy - \sqrt{2}\bar{\theta}\bar{\theta}\theta\psi(x)\right] =$$

$$= -\frac{i}{16}\dot{\theta}^{\alpha}\left[2\sqrt{2}\int G(x,y)\left(\sigma^{\nu}\bar{\theta}\right)_{\alpha}\partial_{\nu}\partial_{\eta}\psi(y)\sigma^{\eta}\bar{\theta} dy - \right.$$

$$\left. -4\sqrt{2}\int \partial_{\mu}G(x,y)\left(\sigma^{\mu}\bar{\theta}\right)_{\alpha}\partial_{\eta}\psi(y)\sigma^{\eta}\bar{\theta} dy - \sqrt{2}\bar{\theta}\bar{\theta}\psi_{\alpha}(x)\right];$$

(2.37)

$$-i\dot{\theta}^{\alpha} D_{\alpha} S^{(A)}(x,\theta,\bar{\theta}) = -\frac{i}{16}\dot{\theta}^{\alpha} D_{\alpha} \int G(x,y) D^{2}\Phi^{(A)}(y,\theta,\bar{\theta}) dy =$$

$$= -\frac{i}{16}\dot{\theta}^{\alpha}\left(\frac{\partial}{\partial\theta^{\alpha}} + i\sigma^{\mu}{}_{\alpha\dot{\beta}}\bar{\theta}^{\dot{\beta}}\partial_{\mu}\right)\left[-4\bar{\theta}\bar{\theta}A(x)\right] = 0.$$

(2.38)

Substituting such obtained expressions (2.34) and (2.37) into (1.29), we achieve the required interaction Lagrangian describing the movement of a superparticle in an external chiral matter superfield

$$L_{\text{int}} = q\left[\frac{1}{4}\left(\dot{x}^{\mu} - i\left[\left(\theta\sigma^{\mu}\dot{\bar{\theta}}\right) - \left(\dot{\theta}\sigma^{\mu}\bar{\theta}\right)\right]\right)\left\{\frac{1}{16}\left[4\sqrt{2}\psi(x)\sigma^{\mu}\bar{\theta} - i\sqrt{2}\theta\theta\partial^{\mu}(\theta\psi(x)) + \right.\right.$$

$$+ i2\sqrt{2}\theta\theta\partial_{\nu}(\theta\sigma^{\nu\mu}\psi(x)) + i4\sqrt{2}(\theta\sigma^{\mu}\bar{\theta})\partial_{\nu}\psi(x)\sigma^{\nu}\bar{\theta} -$$

$$-4\sqrt{2}\left\{\int \partial_{\nu}G(x,y)\partial^{\nu}(\psi(y)\sigma^{\mu}\bar{\theta}) dy + \int \partial^{\mu}G(x,y)\partial_{\nu}(\psi(y)\sigma^{\nu}\bar{\theta}) dy - \right.$$

$$- \int \partial_{\nu}G(x,y)\partial^{\mu}(\psi(y)\sigma^{\nu}\bar{\theta}) dy + i\int \varepsilon^{\nu\mu\lambda\rho}\partial_{\lambda}G(x,y)\partial_{\nu}(\psi(y)\sigma_{\rho}\bar{\theta}) dy\right\} +$$

$$+ i2\sqrt{2}(\theta\sigma^{\nu}\bar{\theta})\left\{\int \partial_{\nu}G(x,y)\partial^{\mu}\partial_{\eta}\psi(y)\sigma^{\eta}\bar{\theta} dy - \int g_{\mu\nu}\partial^{\nu}G(x,y)\partial_{\nu}\partial_{\eta}\psi(y)\sigma^{\eta}\bar{\theta} dy + \right.$$

$$+ \int \partial^{\mu}G(x,y)\partial_{\nu}\partial_{\eta}\psi(y)\sigma^{\eta}\bar{\theta} dy - i\int \varepsilon^{\rho\mu\lambda}{}_{\nu}\partial_{\rho}G(x,y)\partial_{\lambda}\partial_{\eta}\psi(y)\sigma^{\eta}\bar{\theta} dy -$$

$$- 4\int \partial^{\mu}\partial_{\nu}G(x,y)\partial_{\eta}\psi(y)\sigma^{\eta}\bar{\theta} dy\right\}\right] - \ c.c.\} +$$

$$+ \frac{i}{16}\dot{\theta}^{\alpha}\left[2\sqrt{2}\int G(x,y)\left(\sigma^{\nu}\bar{\theta}\right)_{\alpha}\partial_{\nu}\partial_{\eta}\psi(y)\sigma^{\eta}\bar{\theta} dy - \right.$$

$$\left. -4\sqrt{2}\int \partial_{\mu}G(x,y)\left(\sigma^{\mu}\bar{\theta}\right)_{\alpha}\partial_{\eta}\psi(y)\sigma^{\eta}\bar{\theta} dy - \sqrt{2}\bar{\theta}\bar{\theta}\psi_{\alpha}(x)\right] + c.c.\right]. \quad (2.39)$$

## Conclusion

In this paper we have proposed the interaction Lagrangian for a superparticle moving through an external chiral matter superfield that is invariant under the supersymmetry transformation. The resulting expression (2.39) is somewhat cumbersome and non-local. The nonlocality is an inevitable consequence of the relation (2.27) and is the main disadvantage of the proposed approach. To circumvent this problem, we can try to match the external fields in such a way as to get rid of the nonlocality[1]. This will be the subject of further research.

## Appendix
### Useful spinor identities

We accept the notions of Wess and Bagger [19]

$$g_{\mu\nu} = \begin{pmatrix} -1 & 0 & 0 & 0 \\ 0 & 1 & 0 & 0 \\ 0 & 0 & 1 & 0 \\ 0 & 0 & 0 & 1 \end{pmatrix} \quad (A.1)$$

$$\varepsilon_{12} = \varepsilon^{12} = 1, \quad \varepsilon_{12} = \varepsilon^{21} = -1, \quad \varepsilon_{11} = \varepsilon^{22} = 0 \quad (A.2)$$

$$\psi^\alpha = \varepsilon^{\alpha\beta}\psi_\beta, \quad \psi_\alpha = \varepsilon_{\alpha\beta}\psi^\beta \quad (A.3)$$

$$\psi\chi = \psi^\alpha \chi_\alpha = -\psi_\alpha \chi^\alpha = \chi^\alpha \psi_\alpha = \chi\psi$$
$$\bar{\psi}\bar{\chi} = \bar{\psi}_{\dot\alpha}\bar{\chi}^{\dot\alpha} = -\bar{\psi}^{\dot\alpha}\bar{\chi}_{\dot\alpha} = \bar{\chi}_{\dot\alpha}\bar{\psi}^{\dot\alpha} = \bar{\chi}\bar{\psi} \quad (A.4)$$

Bellow we give the form of the $\sigma$-matrices, some spinor algebra and the definitions of the covariant spinor derivatives [19, 20]:

$$\sigma^0 = \begin{pmatrix} -1 & 0 \\ 0 & -1 \end{pmatrix} \quad \sigma^1 = \begin{pmatrix} 0 & 1 \\ 1 & 0 \end{pmatrix}$$
$$\sigma^2 = \begin{pmatrix} 0 & -i \\ i & 0 \end{pmatrix} \quad \sigma^3 = \begin{pmatrix} 1 & 0 \\ 0 & -1 \end{pmatrix} \quad (A.5)$$

---

[1] The idea of getting rid of nonlocality for particular configurations of external fields was expressed in a private conversation by I.L. Buchbinder to one of the authors (Yu.A.M.) of this paper.

$$\bar{\sigma}^{\mu\alpha\beta} = \varepsilon^{\dot{\alpha}\dot{\beta}}\varepsilon^{\alpha\beta}\sigma_{\beta\dot{\beta}}{}^{\mu}$$

$$\bar{\sigma}^0 = \sigma^0$$
$$\bar{\sigma}^{1,2,3} = -\sigma^{1,2,3}$$
(A.6)

$$Tr\bar{\sigma}^{\mu}\sigma^{\nu} = -2g^{\mu\nu}$$
(A.7)

$$\left(\sigma^{\mu}\bar{\sigma}^{\nu} + \sigma^{\nu}\bar{\sigma}^{\mu}\right)_{\alpha}^{\beta} = -2g^{\mu\nu}\delta_{\alpha}^{\beta}$$
$$\left(\bar{\sigma}^{\mu}\sigma^{\nu} + \bar{\sigma}^{\nu}\sigma^{\mu}\right)_{\dot{\beta}}^{\dot{\alpha}} = -2g^{\mu\nu}\delta_{\dot{\beta}}^{\dot{\alpha}}$$
(A.8)

$$\sigma^{\nu\mu\beta}{}_{\alpha} = \frac{1}{4}\left(\sigma_{\alpha\dot{\alpha}}{}^{\nu}\bar{\sigma}^{\mu\dot{\alpha}\beta} - \sigma_{\alpha\dot{\alpha}}{}^{\mu}\bar{\sigma}^{\nu\dot{\alpha}\beta}\right)$$
$$\bar{\sigma}^{\nu\mu\dot{\alpha}}{}_{\dot{\beta}} = \frac{1}{4}\left(\bar{\sigma}^{\nu\dot{\alpha}\alpha}\sigma_{\alpha\dot{\beta}}{}^{\mu} - \bar{\sigma}^{\mu\dot{\alpha}\alpha}\sigma_{\alpha\dot{\beta}}{}^{\nu}\right)$$
(A.9)

$$\sigma^a\bar{\sigma}^b\sigma^c + \sigma^c\bar{\sigma}^b\sigma^a = 2\left(g^{ac}\sigma^b - g^{bc}\sigma^a - g^{ab}\sigma^c\right)$$
$$\bar{\sigma}^a\sigma^b\bar{\sigma}^c + \bar{\sigma}^c\sigma^b\bar{\sigma}^a = 2\left(g^{ac}\bar{\sigma}^b - g^{bc}\bar{\sigma}^a - g^{ab}\bar{\sigma}^c\right)$$
(A.10)

$$\sigma^a\bar{\sigma}^b\sigma^c - \sigma^c\bar{\sigma}^b\sigma^a = 2i\varepsilon^{abcd}\sigma_d$$
$$\bar{\sigma}^a\sigma^b\bar{\sigma}^c - \bar{\sigma}^c\sigma^b\bar{\sigma}^a = -2i\varepsilon^{abcd}\bar{\sigma}_d$$
(A.11)

$$\theta^{\alpha}\theta^{\beta} = -\frac{1}{2}\varepsilon^{\alpha\beta}\theta\theta \qquad \theta_{\alpha}\theta_{\beta} = \frac{1}{2}\varepsilon_{\alpha\beta}\theta\theta$$
$$\bar{\theta}^{\dot{\alpha}}\bar{\theta}^{\dot{\beta}} = \frac{1}{2}\varepsilon^{\dot{\alpha}\dot{\beta}}\bar{\theta}\bar{\theta} \qquad \bar{\theta}_{\dot{\alpha}}\bar{\theta}_{\dot{\beta}} = -\frac{1}{2}\varepsilon_{\dot{\alpha}\dot{\beta}}\bar{\theta}\bar{\theta}$$
(A.12)

$$\frac{\partial}{\partial\theta_{\alpha}} = -\varepsilon^{\alpha\beta}\frac{\partial}{\partial\theta^{\beta}} \qquad \frac{\partial}{\partial\bar{\theta}_{\dot{\alpha}}} = -\varepsilon^{\dot{\alpha}\dot{\beta}}\frac{\partial}{\partial\bar{\theta}^{\dot{\beta}}}$$
(A.13)

$$\frac{\partial}{\partial\theta^{\alpha}}\theta^{\beta} = \frac{\partial}{\partial\theta_{\alpha}}\theta_{\beta} = \delta_{\alpha}^{\beta} \qquad \frac{\partial}{\partial\theta_{\alpha}}\theta^{\beta} = -\varepsilon^{\alpha\beta} \qquad \frac{\partial}{\partial\theta^{\alpha}}\theta_{\beta} = -\varepsilon_{\alpha\beta}$$
$$\frac{\partial}{\partial\bar{\theta}_{\dot{\alpha}}}\bar{\theta}_{\dot{\beta}} = \frac{\partial}{\partial\bar{\theta}^{\dot{\alpha}}}\bar{\theta}^{\dot{\beta}} = \delta_{\dot{\alpha}}^{\dot{\beta}} \qquad \frac{\partial}{\partial\bar{\theta}_{\dot{\alpha}}}\bar{\theta}^{\dot{\beta}} = -\varepsilon^{\dot{\alpha}\dot{\beta}} \qquad \frac{\partial}{\partial\bar{\theta}^{\dot{\alpha}}}\bar{\theta}_{\dot{\beta}} = -\varepsilon_{\dot{\alpha}\dot{\beta}}$$
(A.14)

$$\frac{\partial}{\partial\theta^{\alpha}}\theta\theta = 2\theta_{\alpha} \qquad \frac{\partial}{\partial\theta_{\alpha}}\theta\theta = -2\theta^{\alpha}$$
$$\frac{\partial}{\partial\bar{\theta}_{\dot{\alpha}}}\bar{\theta}\bar{\theta} = 2\bar{\theta}^{\dot{\alpha}} \qquad \frac{\partial}{\partial\bar{\theta}^{\dot{\alpha}}}\bar{\theta}\bar{\theta} = -2\bar{\theta}_{\dot{\alpha}}$$
(A.15)

$$D_\alpha = \frac{\partial}{\partial \theta^\alpha} + i\sigma^\mu{}_{\alpha\dot\beta}\bar\theta^{\dot\beta}\partial_\mu \qquad D^\alpha = -\frac{\partial}{\partial \theta_\alpha} - i\bar\theta_{\dot\beta}\bar\sigma^{\mu\alpha\dot\beta}\partial_\mu$$

$$\bar D_{\dot\alpha} = -\frac{\partial}{\partial \bar\theta^{\dot\alpha}} - i\theta^\beta \sigma^\mu{}_{\beta\dot\alpha}\partial_\mu \qquad \bar D^{\dot\alpha} = \frac{\partial}{\partial \bar\theta_{\dot\alpha}} + i\bar\sigma^{\mu\dot\alpha\beta}\theta_\beta \partial_\mu \qquad (A.16)$$